\documentstyle[prb,aps,multicol,eqsecnum,epsfig]{revtex}

\begin{document}
\title{A tetragonal-to-monoclinic phase transition in a ferroelectric perovskite:
the structure of PbZr$_{0.52}$Ti$_{0.48}$O$_{3}$.}
\author{B. Noheda\thanks{%
Visiting scientist at Brookhaven National Laboratory.}, J. A. Gonzalo}
\address{Dep. Fisica de Materiales, UAM. Cantoblanco. 28049-Madrid. Spain }
\author{L.E. Cross, R. Guo, S-E. Park}
\address{Mat. Res. Lab., The Pennsylvania State University, PA 16802-4800}
\author{D.E. Cox and G. Shirane.}
\address{Physics Department. Brookhaven National Laboratory. Upton, NY 11973-5000}
\date{\today }
\maketitle

\begin{abstract}
The perovskite-like ferroelectric system PbZr$_{1-x}$Ti$_{x}$O$_{3}$ (PZT)
has a nearly vertical morphotropic phase boundary (MPB) around x= 0.45-0.50.
Recent synchrotron x-ray powder diffraction measurements by Noheda {\it et
al.} $\left[ \text{Appl. Phys. Lett.{\bf \ 74}, 2059 (1999)}\right] $ have
revealed a new monoclinic phase between the previously-established
tetragonal and rhombohedral regions. In the present work we describe a
Rietveld analysis of the detailed structure of the tetragonal and monoclinic
PZT phases on a sample with x= 0.48 for which the lattice parameters are
respectively: a$_{t}$= 4.044 \AA , c$_{t}$= 4.138 \AA , at 325 K, and a$_{m}$%
= 5.721 \AA , b$_{m}$= 5.708 \AA , c$_{m}$= 4.138 \AA , $\beta $= 90.496$^{o}
$, at 20K. In the tetragonal phase the shifts of the atoms along the polar $%
[001]$ direction are similar to those in PbTiO$_{3}$ but the refinement
indicates that there are, in addition, local disordered shifts of the Pb
atoms of $\sim $0.2 \AA\  perpendicular to the polar axis. The monoclinic structure
can be viewed as a condensation along one of the $\langle $110$\rangle $
directions of the local displacements present in the tetragonal phase. It
equally well corresponds to a freezing-out of the local displacements along
one of the $\langle $100$\rangle $ directions recently reported by Corker 
{\it et al. }$\left[ \text{J. Phys. Condens. Matter {\bf 10}, 6251 (1998)}%
\right] $ for rhombohedral PZT. The monoclinic structure therefore provides
a microscopic picture of the MPB region in which one of the ``locally''
monoclinic phases in the ``average'' rhombohedral or tetragonal structures
freezes out, and thus represents a bridge between these two phases.
\end{abstract}


\preprint{cond-mat/123-ms}

\section{Introduction}

\label{sec:level1}

Perovskite-like oxides have been at the center of research on ferroelectric
and piezoelectric materials for the past fifty years because of their simple
cubic structure at high temperatures and the variety of high symmetry phases
with polar states found at lower temperatures. Among these materials the
ferroelectric PbZr$_{1-x}$Ti$_{x}$O$_{3}$ (PZT) solid solutions have
attracted special attention since they exhibit an unusual phase boundary
which divides regions with rhombohedral and tetragonal structures, called
the ``morphotropic phase boundary'' (MPB) by Jaffe {\it et al.}\cite{Jaffe}.
Materials in this region exhibit a very high piezoelectric response, and it
has been conjectured that these two features are intrinsically related. The
simplicity of the perovskite structure is in part responsible for the
considerable progress made recently in the determination of the basic
structural properties and stability of phases of some important perovskite
oxides, based on {\it ab-initio} calculations (see, e.g., \cite
{Cohen,King-Smith,Zhong1,Zhong2,A. Garcia,Rabe,Cockayne,Ghosez}). Recently,
such calculations have also been used to investigate solid solutions and, in
particular, PZT, where the effective Hamiltonian includes both structural
and compositional degrees of freedom \cite{Bellaiche1,Bellaiche2,Saghi} .

The PZT phase diagram of Jaffe {\it et al.} \cite{Jaffe}, which covers only
temperatures above 300 K, has been accepted as the basic characterization of
the PZT solid solution. The ferroelectric region of the phase diagram
consists mainly of two different regions: the Zr-rich rhombohedral region, (F%
$_{R}$) that contains two phases with space groups R3m and R3c, and the
Ti-rich tetragonal region (F$_{T}$), with space group P4mm\cite{Shirane}.
The two regions are separated by a boundary that is nearly independent of
temperature, the MPB mentioned above, which lies at a composition close to
x= 0.47. Many structural studies have been reported around the MPB, since
the early 1950's, when these solid solutions were first studied\cite
{Shirane,Sawaguchi}, since the high piezoelectric figure-of-merit that makes
PZT so extraordinary is closely associated with this line \cite{Jaffe,Xu}
.The difficulty in obtaining good single crystals in this region, and the
characteristics of the boundary itself, make good compositional homogeneity
essential if single phase ceramic materials are to be obtained. Because of
this, the MPB is frequently reported as a region of phase-coexistence whose
width depends on the sample processing conditions\cite
{Kakewaga,Fernandes,Hammer,Cao}.

Recently, an entirely new feature of the morphotropic phase boundary has
been revealed by the discovery of a new ferroelectric monoclinic phase (F$%
_{M}$) in the Pb(Zr$_{1-x}$Ti$_{x}$)O$_{3}$ ceramic system\cite{Noheda}.
From a synchrotron x-ray powder diffraction study of a composition with
x=0.48, a tetragonal-to-monoclinic phase transition was discovered at $\sim $%
300 K. The monoclinic unit cell is such that a$_{m}$ and b$_{m}$ lay along
the tetragonal $[\overline{1}\overline{1}0]$ and $[1\overline{1}0]$
directions (a$_{m}\approx $ b$_{m}\approx $ a$_{t}\surd $2), and c$_{m}$
deviates slightly from the [001] direction (c$_{m}\approx $ c$_{t}$)\cite
{footnote} . The space group is Cm, and the temperature dependence of the
monoclinic angle $\beta $ gives immediately the evolution of the order
parameter for the tetragonal-monoclinic (F$_{T}$-F$_{M}$) transition. The
polar axis of the monoclinic cell can in principle be directed along any
direction within the {\it ac} mirror plane, making necessary a detailed
structural study to determine its direction.

In the present work we present such a detailed structure determination of
the monoclinic phase at 20 K and the tetragonal phase at 325 K in PZT with
x= 0.48. The results show that the polarization in the monoclinic plane lies
along a direction between the pseudo-cubic [001]$_{c}$ and [111]$_{c}$
directions, corresponding to the first example of a species with P$_{x}^{2}$%
= P$_{y}^{2}\neq $ P$_{z}^{2}$. A tentative phase diagram is presented in
Figure 1, which includes data for the x= 0.48 composition together with
those of the recently-studied x= 0.50 composition \cite{Noheda2}. The most
striking finding, however, is that the monoclinic cation displacements found
here correspond to one of the three locally disordered sites reported by
Corker {\it et al.} \cite{Corker} for rhombohedral compositions in the
region x= 0.1-0.4, and thus provide a microscopic model of the
rhombohedral-to-monoclinic phase transition. This, together with the fact
that the space group of the new phase, Cm, is a subgroup of both P4mm and
R3m, suggests that F$_{M}$ represents an intermediate phase connecting the
well-known F$_{T}$ and F$_{R}$ PZT phases.

\section{Experimental\label{sec:level2}\protect\bigskip}

A PZT sample with x= 0.48 was prepared by conventional solid-state reaction
techniques using appropriate amounts of reagent-grade powders of lead
carbonate, zirconium oxide and titanium oxide, with chemical purities better
than 99.9$\%$. Pellets were pressed and heated to 1250$^{o}$C at a ramp rate
of 10$^{o}$C/min, held at this temperature in a covered crucible for 2
hours, and furnace-cooled. During sintering, PbZrO$_{3}$ was used as a lead
source in the crucible to minimize volatilization of lead.

High-resolution synchrotron x-ray powder diffraction measurements were made
at beam line X7A at the Brookhaven National Synchrotron Light Source. In the
first set of measurements, an incident beam of wavelength 0.6896 \AA\ from a
Ge(111) double-crystal monochromator was used in combination with a Ge(220)
crystal and scintillation detector in the diffraction path. The resulting
instrumental resolution is about 0.01$^{o}$ on the 2$\theta $ scale, an
order-of-magnitude better than that of a laboratory instrument. Data were
collected from a disk in symmetric flat-plate reflection geometry over
selected angular regions in the temperature range 20-736 K. Coupled $\theta $%
- 2$\theta $ scans were performed over selected angular regions with a 2$%
\theta $ step interval of 0.01$^{o}$. The sample was rocked 1-2$^{o}$ during
data collection to improve powder averaging.

Measurements above room temperature were performed with the disk mounted on
a BN sample pedestal inside a wire-wound BN tube furnace. The furnace
temperature was measured with a thermocouple mounted just below the pedestal
and the temperature scale calibrated with a sample of CaF$_{2}$. The
accuracy of the temperature in the furnace is estimated to be within 10 K,
and the temperature stability about 2 K. For low temperature measurements,
the pellet was mounted on a Cu sample pedestal and loaded in a closed-cycle
He cryostat, which has an estimated temperature accuracy of 2 K and
stability better than 0.1 K. The diffracted intensities were normalized with
respect to the incident beam monitor.

For the second set of measurements aimed at the detailed determination of
the structure, a linear position-sensitive detector was mounted on the 2$%
\theta $ arm of the diffractometer instead of the crystal analyzer, and a
wavelength of 0.7062 \AA\ was used. This configuration gives
greatly-enhanced counting rates which make it feasible to collect accurate
data from very narrow-diameter capillary samples in Debye-Scherrer geometry,
with the advantage that systematic errors due to preferred orientation or
texture effects are largely eliminated. A small piece of the sintered disk
was carefully crushed and sealed into a 0.2 mm diameter glass capillary. The
latter was loaded into a closed-cycle cryostat, and extended data sets were
collected at 20 and 325 K while the sample was rocked over a 10$^{o}$\
range. With this geometry the instrumental resolution is about 0.03$^{o}$ on
the 2$\theta $ scale. Because lead is highly absorbing, the data were
corrected for absorption effects\cite{Dwiggins} based on an approximate
value of $\mu r=1.4$ determined from the weight and dimensions of the sample.

\bigskip

\section{Phase transitions}

\label{sec:level3}The evolution of the lattice parameters with temperature
was briefly summarized in ref.[20], and a more complete analysis is
presented below.{\bf \ }The results of the full structure analysis are
described later.

A transition from the cubic to the tetragonal phase was observed at $\sim $%
660 K, in agreement with the phase diagram shown in Fig. 1. The measurements
made on the pellet in the cubic phase at 736 K demonstrate the excellent
quality of the sample, which exhibits diffraction peaks with full-widths at
half-maximum (FWHM) ranging from 0.01 to 0.03$^{o}$ as shown for the (111)
reflection plotted as the upper-right inset in Fig. 2. The FWHM's ($\Gamma $%
) for several peaks were determined from least-squares fits to a
pseudo-Voigt function with the appropriate corrections for asymmetry effects 
\cite{Finger}, and corrected for instrumental resolution. The corrected
values are shown in Fig. 2 in the form of a Williamson-Hall plot \cite
{Williamson-Hall}

$\label{form}$%
\begin{equation}
\Gamma \cos \theta =\lambda /L+2(\Delta d/d)\sin \theta
\end{equation}
where $\lambda $ is the wavelength and $L$ is the mean crystallite size.
From the slope of a linear fit to the data, the distribution of d-spacings, $%
\Delta $d/d, is estimated to be $\sim 3\times 10^{-4}$, corresponding to a
compositional inhomogeneity $\Delta $x of less than $\pm $0.003{\bf .} From
the intercept of the line on the ordinate axis the mean crystallite size is
estimated to be $\sim $1 $\mu $m.

A tetragonal-to-monoclinic phase transition in PZT with x= 0.48 was recently
reported by Noheda {\it et al.}\cite{Noheda}. Additional data have been
obtained near the phase transition around 300 K which have allowed a better
determination of the phase transition to be made, as shown by the evolution
of the lattice parameters as a function of temperature in Figure 3. The
tetragonal strain, c$_{t}$/a$_{t}$, increases as the temperature decreases
from the Curie point (T$\approx $ 660K), to a value of 1.0247 at 300 K,
below which peak splittings characteristic of a monoclinic phase with a$%
_{m}\approx $ b$_{m}\approx $ a$_{t}\surd 2$ , $\beta \neq 90^{o}$, are
observed (Figure 3). As the temperature continues to decrease down to 20 K, a%
$_{m}$ (which is defined to lie along the [$\overline{1}\overline{1}$0]
tetragonal direction) increases very slightly, and b$_{m}$ (which lies along
the [1$\overline{1}$0] tetragonal direction) decreases. The c$_{m}$ lattice
parameter reaches a broad maximum value of 4.144 \AA\ between 240-210 K and
then reaches a shallow minimum value of 4.137 \AA\ at 60 K.\ Over the same
temperature region there is a striking variation of $\Delta $d/d determined
from Williamson-Hall plots at various temperatures, as shown in the
upper-left inset in Fig. 2. $\Delta $d/d increases rapidly \ as the
temperature approaches the F$_{T}$-F$_{M}$ transition at 300K, in a similar
fashion to the tetragonal strain, and then decreases rapidly below this
temperature in the monoclinic region. Thus the microstrain responsible for
the large increase in $\Delta $d/d is an important feature of the phase
transition, which may be associated with the development of local monoclinic order, and 
is very likely responsible for the large electromechanical response of PZT close to the MPB \cite{Jaffe}.

The deviation of the monoclinic angle $\beta $ from 90$^{o}$ is an order
parameter of the F$_{T}$-F$_{M}$ transition, and its evolution with
temperature is also depicted in Figure 3. This phase transition presents a
special problem due to the steepness of the phase boundary (the MPB in Fig.
1). As shown in the previous section, the compositional fluctuations are
quite small in these ceramic samples ($\Delta $x$\approx $ $\pm $0.003) but,
even in this case, the nature of the MPB implies an associated temperature
uncertainty of $\Delta $T$\approx $ 100 K. There is, therefore, a rather
wide range of transition temperatures instead of a single well-defined
transition, so that the order parameter is smeared out as a function of
temperature around the phase change, thereby concealing the nature of the
transition.

Scans over the (220)$_{c}$ region for several different temperatures are
plotted in Figure 4, which shows the evolution of phases from the cubic
phase at 687 K (upper-left plot) to the monoclinic phase at 20 K
(lower-right plot), passing through the tetragonal phase at intermediate
temperatures. With decreasing temperature, the tetragonal phase appears at $%
\sim $660 K and the development of the tetragonal distortion can be observed
on the left side of the figure from the splitting of the (202)$_{t}$ and
(220)$_{t}$ reflections. On the right side of the figure, the evolution of
the monoclinic phase, which appears below $\sim $300 K, is shown by the
splitting into the (22$\overline{2}$)$_{m}$, (222)$_{m}$, (400)$_{m}$ and
(040)$_{m}$ monoclinic reflections. It is quite evident from Fig. 4 that the
(202)$_{t}$ peak is much broader than the neighboring (220)$_{t}$ peak, for
example, and this ``anisotropic'' peak broadening is a general feature of
the diffraction data for both phases. Another feature of the patterns is the
presence of additional diffuse scattering between neighboring peaks, which
is particularly evident between tetragonal (00l) and (h00) pairs, and the
corresponding monoclinic (00l) and (hh0) pairs.

\section{Structure determination}

A detailed analysis of the 325 K tetragonal and 20 K monoclinic structures
of PbZr$_{0.52}$Ti$_{0.48}$O$_{3}$ was carried out by Rietveld refinement
using the GSAS program package \cite{GSAS}. The pseudo-Voigt peak shape
function option was chosen\cite{Finger} and background was estimated by
linear interpolation between fixed values. An important feature of the
refinements was the need to allow for the anisotropic peak broadening
mentioned above. This was accomplished by the use of the
recently-incorporated generalized model for anisotropic peak broadening
proposed by Stephens\cite{Stephens}, which is based on a distribution of
lattice parameters. It was also necessary to take into account some
additional diffuse scattering by modeling with a second, cubic, phase with
broad, predominately Gaussian, peaks. A similar strategy has been adopted by
Muller {\it et al.}\cite{Muller} in a recent study of PbHf$_{0.4}$Ti$_{0.6}$O%
$_{3}$. Although in principle this could represent a fraction of
untransformed cubic phase, we suspect that the diffuse scattering is
associated with locally disordered regions in the vicinity of domain walls.
The refinements were carried out with the atoms assigned fully-ionized
scattering factors.

\subsection{\protect\bigskip Tetragonal structure at 325 K}

At 325 K the data show tetragonal symmetry similar to that of PbTiO$_{3}$.
This tetragonal structure has the space group P4mm with Pb in 1(a) sites at
(0, 0, z); Zr/Ti and O(1) in 1(b) sites at (1/2, 1/2, z) and O$(2)$ in 2(c)
sites at (1/2, 0, z). For the refinement we adopt the same convention as
that used in refs.[30] and [31] for PbTiO$_{3}$, with Pb fixed at (0,0,0).
However, instead of thinking in terms of shifts of the other atoms with
respect to this origin, it is more physically intuitive to consider
displacements of Pb and Zr/Ti from the center of the distorted oxygen
cuboctahedra and octahedra, respectively. We shall take this approach in the
subsequent discussion.

The refinement was first carried out with individual isotropic (U$_{iso}$)
temperature factors assigned. Although a reasonably satisfactory fit was
obtained (R$_{F^{2}}$= 8.9\%), U$_{iso}$ for O(1) was slightly negative and U%
$_{iso}$ for Pb was very large, 0.026 \AA $^{2}$, much larger than U$_{iso}$
for the other atoms. Similarly high values for Pb(U$_{iso}$) in Pb-based
perovskites are well-known in the literature, and are usually ascribed to
local disordered displacements, which may be either static or dynamic. Refinement with anisotropic temperature factors 
\cite{footnote2} (U$_{11}$ and U$_{33}$) assigned to Pb (Table I, model I)
gave an improved fit (R$_{F^{2}}$= 6.1\%) with U$_{11}$(= U$_{22}$)
considerably larger than U$_{33}$ (0.032 and 0.013 \AA $^{2}$, respectively)
corresponding to large displacements perpendicular to the polar [001] axis.
A further refinement based on local displacements of the Pb from the 1(a)
site to the 4(d) sites at (x,x,0), with isotropic temperature factors
assigned to all the atoms{\bf , }gave{\bf \ }a small improvement in the fit
( R$_{F^{2}}$= 6.0\%) with x$\simeq 0.033$, corresponding to local shifts
along the $\langle $110$\rangle $ axes, and a much more reasonable
temperature factor (Table I, model II). In order to check that high correlations
 between the temperature factor and local displacements were not biasing the result 
of this refinement, we have applied a commonly-used procedure consisting of a series 
of refinements based on Model II in which Pb displacements along $\langle $110$\rangle $ 
were fixed but all the other parameters were varied \cite{Itoh,Malibert}. Fig. 5 shows
unambiguously that there is well-defined minimum in the R-factor for a 
displacement of about 0.19 \AA, consistent with the result in Table I.
A similar minimum was obtained for shifts along $\langle $100$\rangle $ directions with a 
slightly higher R-factor. Thus, in addition to a shift of 0.48
\AA\ for Pb along the polar [001] axis towards four of its O(2) neighbors,
similar to that in PbTiO$_{3}$\cite{Glazer&Mabud,Nelmes,Shirane&Pepinski},
there is a strong indication of substantial local shifts of $\sim $0.2 \AA\
perpendicular to this axis. The Zr/Ti displacement is 0.27 \AA\ along the polar axis, 
once again similar to the Ti shift in PbTiO$_{3}$. Attempts to model local displacements along $\langle $110$\rangle $ directions for the Zr/Ti atoms were unsuccessful due
to the large correlations between these shifts and the temperature factor.
Further attempts to refine the z parameters of the Zr and Ti atoms
independently, as Corker {\it et al.} were able to do\cite{Corker}, were
likewise unsuccessful, presumably because the scattering contrast for x-rays
is much less than for neutrons.

From the values of the atomic coordinates listed in Table \ref{Table1}, it
can be inferred that the oxygen octahedra are somewhat more distorted than
in PbTiO$_{3}$, the O(1) atoms being displaced 0.08 \AA\ towards the O(2)
plane above. The cation displacements are slightly larger than those
recently reported by Wilkinson {\it et al.} \cite{Wilkinson} for samples
close to the MPB containing a mixture of rhombohedral and tetragonal phases, and 
in excellent agreement with the theoretical values obtained by L. Bellaiche and D. Vanderbilt \cite{Bellaiche3} for PZT with x= 0.50 from first principles calculations.
As far as we are aware no other structural analysis of PZT compositions in
the tetragonal region has been reported in the literature.

Selected bond distances for the two models are shown in Table \ref{Table2}.
For model I, Zr/Ti has short and long bonds with O(1) of 1.85 and 2.29 \AA ,
respectively, and four intermediate-length O(2) bonds of 2.05 \AA . There
are four intermediate-length Pb-O(1) bonds \ of 2.89 \AA , four short
Pb-O(2) bonds of 2.56 \AA\ and four much larger Pb-O(2) distances of 3.27\
\AA . For model II, the Zr/Ti-O distances are the same, but the Pb-O
distances change significantly. A Pb atom in one of the four equivalent
(x,x,0) sites in Table \ref{Table1} now has a highly distorted coordination,
consisting of two short and two intermediate Pb-O(2) bonds of 2.46 and 2.67
\AA , and one slightly longer Pb-O(1) bond of 2.71 \AA\ (Table \ref{Table2}%
). The tendency of Pb$^{+2}$, which has a{\bf \ }lone{\bf \ }sp electron
pair, to form short covalent bonds with a few neighboring oxygens is well
documented in the literature\cite{Corker,Teslic,Corker2,Teslic2}.

The observed and calculated diffraction profiles and the difference plot are
shown in Figure \ref{fig6} for a selected $2\theta $ range between 7 and 34$%
^{\circ }$ (upper figure). The short vertical markers represent the
calculated peak positions. The upper and lower sets of markers correspond to
the cubic and tetragonal phases, respectively. We note that although
agreement between the observed and the calculated profiles is considerably
better when the diffuse scattering is modeled with a cubic phase, the
refined values of the atomic coordinates are not significantly affected by
the inclusion of this phase. The anisotropic peak broadening was found to be
satisfactorily described by two of the four parameters in the generalized
model for tetragonal asymmetry\cite{Stephens}.

\subsection{\protect\bigskip Monoclinic structure at 20 K}

As discussed above, the diffraction data at 20 K can be indexed
unambiguously \ on the basis of a monoclinic cell with space group Cm. In
this case Pb, Zr/Ti and O(1) are in 2(a) sites at (x, 0, z), and O(2) in
4(b) sites at (x, y, z). Individual isotropic temperature factors were
assigned, and Pb was fixed at (0,0,0). For monoclinic symmetry, the
generalized expression for anisotropic peak broadening contains nine
parameters, but when all of these were allowed to vary the refinement was
slightly unstable and did not completely converge. After several tests in
which some of the less significant values were fixed at zero, satisfactory
convergence was obtained with 3 parameters ($R_{wp}=0.036$ , $\chi ^{2}=11.5$%
). During these tests, there was essentially no change in the refined values
of the atomic coordinates. A small improvement in the fit was obtained when
anisotropic temperature factors were assigned to Pb ($R_{wp}=0.033,$ $\chi
^{2}=9.2$). The final results are listed in Table \ref{Table3}, and the
profile fit and difference plot are shown in the lower part of Fig. \ref
{fig6}.

From an inspection of the results in Tables \ref{Table1} and \ref{Table3},
it can be seen that the displacements of the Pb and Zr/Ti atoms along [001]
are very similar to those in the tetragonal phase at 325 K, about 0.53 and
0.24\AA , respectively. However, in the monoclinic phase at 20 K, there are
also significant shifts of these atoms along the monoclinic [$\overline{1}$%
00], i.e. pseudo-cubic $[110]$, towards their O(2) neighbors in adjacent
pseudo-cubic (110) planes, of about 0.24 and 0.11 \AA , respectively, which
corresponds to a rotation of the polar axis from [001] towards pseudo-cubic
[111]. The Pb shifts are also qualitatively consistent with the local shifts
of Pb along the tetragonal $\langle 110\rangle $ axes inferred from the
results of model II in Table \ref{Table1}, i.e. about 0.2 \AA . Thus it
seems very plausible that the monoclinic phase results from the condensation
of the local Pb displacements in the tetragonal phase along one of the $%
\langle $110$\rangle $ directions.

Some selected bond distances are listed in Table \ref{Table2}. The
Zr/Ti-O(1) distances are much the same as in the tetragonal structure, but
the two sets of Zr/Ti-O(2) distances are significantly different, 1.96 and
2.13 \AA , compared to the single set at 2.04 \AA\ in the tetragonal
structure. Except for a shortening in the Pb-O(1) distance from 2.71 to 2.60
\AA , the Pb environment is quite similar to that in the tetragonal phase,
with two close O(2) neighbors at 2.46 \AA , and two at 2.64 \AA .

\section{Discussion}

\label{sec:level5}

In the previous section, we have shown that the low-temperature monoclinic
structure of PbZr$_{0.52}$Ti$_{0.48}$O$_{3}$ is derived from the tetragonal
structure by shifts of the Pb and Zr/Ti atoms along the tetragonal [110]
axis. We attribute this phase transition to the condensation of local $%
\langle $110$\rangle $ shifts of Pb which are present in the tetragonal
phase along one of the four $\langle $110$\rangle $ directions. In the
context of this monoclinic structure it is instructive to consider the
structural model for rhombohedral PZT compositions with x= 0.08-0.38
recently reported by Corker {\it el al.}\cite{Corker} on the basis of
neutron powder diffraction data collected at room temperature. In this study
and also an earlier study\cite{Glazer} of a sample with x= 0.1, it was found
that satisfactory refinements could only be achieved with anisotropic
temperature factors, and that the thermal ellipsoid for Pb had the form of a
disk perpendicular to the pseudo-cubic [111] axis. This highly unrealistic
situation led them to a physically much more plausible model involving local
displacements for the Pb atoms of about 0.25 \AA\ perpendicular to the [111]
axis and a much smaller and more isotropic thermal ellipsoid. Evidence for
local shifts of Pb atoms in PZT ceramics has also been demonstrated by
pair-distribution function analysis by Teslic and coworkers\cite{Teslic}.

We now consider the refined values of the Pb atom positions with local
displacements for rhombohedral PZT listed in Table 4 of ref.[23].
With the use of the appropriate transformation matrices, it is
straightforward to show that these shifts correspond to displacements of
0.2-0.25 \AA\ along the direction of the monoclinic [100] axis, similar to
what is actually observed for x= 0.48. It thus seems equally plausible that
the monoclinic phase can also result from the condensation of local
displacements perpendicular to the [111] axis.

The monoclinic structure can thus be pictured as providing a ``bridge''
between the rhombohedral and tetragonal structures in the region of the MPB.
This is illustrated in Table \ref{Table4}, which compares the results for
PZT with x=0.48 obtained in the present study with earlier results\cite{Amin}
for rhombohedral PZT with x= 0.40 expressed in terms of the
monoclinic cell\cite{footnote3}. For x= 0.48, the atomic coordinates for
Zr/Ti, O(1) and O(2) are listed for the ``ideal'' tetragonal structure
(model I) and for a similar structure with local shifts of (0.03, 0.03, 0)
in the first two columns, and for the monoclinic structure in the third
column. The last two columns describe the rhombohedral structure for x= 0.40
assuming local shifts of (-0.02, 0.02, 0) along the hexagonal axes and the
as-refined ``ideal'' structure, respectively. It is clear that the
condensation of these local shifts gives a very plausible description of the
monoclinic structure in both cases. It is also interesting to note the
behavior of the corresponding lattice parameters; metrically the monoclinic
cell is very similar to the tetragonal cell except for the monoclinic angle,
which is close to that of the rhombohedral cell.

Evidence for a tetragonal-to-monoclinic transition in the ferroelectric
material PbFe$_{0.5}$Nb$_{0.5}$O$_{3}$ has also been reported by Bonny {\it %
et al.}\cite{Bonny} from single crystal and synchrotron x-ray powder
diffraction measurements. The latter data show a cubic-tetragonal transition
at $\sim $376K, and a second transition at $\sim $355K. Although the
resolution was not sufficient to reveal any systematic splitting of the
peaks, it was concluded that the data were consistent with a very weak
monoclinic distortion of the pseudo-rhombohedral unit cell. In a recent
neutron and x-ray powder study, Lampis {\it et al.} \cite{Lampis} have shown
that Rietveld refinement gives better agreement for the monoclinic structure
at 80 and 250 K than for the rhombohedral one. The resulting monoclinic
distortion is very weak, and the large thermal factor obtained for Pb is
indicative of a high degree of disorder.

The relationships between the PZT rhombohedral, tetragonal and monoclinic
structures are also shown schematically in Figure 7, in which the
displacements of the Pb atom are shown projected on the pseudo-cubic (110)
mirror plane. The four locally disordered $\langle 110\rangle $ shifts
postulated in the present paper for the tetragonal phase are shown
superimposed on the [001] shift at the left (Fig.7a) and the three locally
disordered $\langle 100\rangle $ shifts proposed by Corker {\it et al.}\cite
{Corker} for the rhombohedral phase are shown superimposed on the [111]
shift at the right (Fig. 7c). It can be seen that both the condensation of
the [110] shift in the tetragonal phase and the condensation of the [001]
shift in the rhombohedral phase leads to the observed monoclinic shift shown
at the center (Fig. 7b). We note that although Corker {\it et al.} discuss
their results in terms of $\langle 100\rangle $ shifts and a [111] shift
smaller than that predicted in the usual refinement procedure, they can be
equally well described by a combination of shifts perpendicular to the [111]
axis and the [111] shift actually obtained in the refinement, as is evident
from Fig. 7c.

We conclude, therefore, that the F$_{M}$ phase establishes a connection
between the PZT phases at both sides of the MPB through the common symmetry
element, the mirror plane, and suggest that there is not really a
morphotropic phase boundary, but rather a ``morphotropic phase'', connecting
the F$_{T}$ and F$_{R}$ phases of PZT.

In the monoclinic phase the difference vector between the positive and
negative centers of charge defines the polar axis, whose orientation, in
contrast to the case of the F$_{T}$ and F$_{R}$ phases, cannot be determined
on symmetry grounds alone. According to this, from the results shown in
Table \ref{Table3}, the polar axis in the monoclinic phase is found to be
tilted about 24$^{o}$ from the [001] axis towards the [111] axis. This
structure represents the first example of a ferroelectric material with P$%
_{x}^{2}$= P$_{y}^{2}$ $\neq $P$_{z}^{2}$, (P$_{x}$, P$_{y}$, P$_{z}$) being
the Cartesian components of the polarization vector. This class corresponds
to the so-called m3m(12)A2Fm type predicted by Shuvalov \cite{Shuvalov}. It
is possible that this new phase is one of the rare examples of a
two-dimensional ferroelectric \cite{Abrahams} in which the unit cell dipole
moment switches within a plane containing the polar axis, upon application
of an electric field.

This new F$_{M}$ phase has important implications; for example, it might
explain the well-known shifts of the anomalies of many physical properties
with respect to the MPB and thus help in understanding the physical
properties in this region, of great interest from the applications
point-of-view\cite{Jaffe}. It has been found that the maximum values of d$%
_{33}$ for rhombohedral PZT with x= 0.40 are not obtained for samples
polarized along the [111] direction but along a direction close to the
perovskite [001] direction \cite{Xiao-hong Du}. This points out the
intrinsic importance of the [001] direction in perovskites, whatever the
distortion present, and is consistent with Corker {\it et al.}'s model for
the rhombohedral phase\cite{Corker}, and the idea of the
rhombohedral-tetragonal transition through a monoclinic phase.

It is also to be expected that other systems with morphotropic phase
boundaries between two non-symmetry-related phases (e.g., other perovskites
or tungsten-bronze mixed systems) may show similar intermediate phases. In
fact, an indication of symmetry lowering at the MPB of the PZN-PT system has
been recently reported by Fujishiro {\it et al.} \cite{Fujishiro}. From a
different point-of-view, a monoclinic ferroelectric perovskite also
represents a new challenge for first-principles theorists, until now used to
dealing only with tetragonal, rhombohedral and orthorhombic perovskites.

A structural analysis of several other PZT compositions with x= 0.42-0.51 is
currently in progress in order to determine the new PZT phase diagram more
precisely. In the preliminary version shown in Fig. 1 we have included data
for a sample with x= 0.50 made under slightly different conditions\cite
{Noheda2} at the Institute of Ceramic and Glass (ICG) in Madrid, together
with the data described in the present work for a sample with x= 0.48
synthesized in the Materials Research Laboratory at the Pennsylvania State
University (PSU). As can be seen the results for these two compositions show
consistent behavior, and demonstrate that the F$_{M}$-F$_{T}$ phase boundary
lies along Jaffe {\it et al.}'s MPB. Preliminary results for a sample from
PSU with x= 0.47 show unequivocally that the monoclinic features are present
at 300 K. However measurements on an ICG sample with the same nominal
composition do not show monoclinicity unambiguously, but instead a rather
complex poorly-defined region from 300-400K between the rhombohedral and
tetragonal phases\cite{Noheda2}. The extension of the monoclinic region and
the location of the F$_{R}$-F$_{M}$ phase boundary are still somewhat
undefined, although it is clear that the monoclinic region has a narrower
composition range as the temperature increases. The existence of a quadruple
point in the PZT phase diagram is an interesting possibility.

Acknowledgments

We wish to gratefully acknowledge B. Jones for the excellent quality of the
x= 0.48 sample used in this work, and we thank A. M. Glazer, E. Moshopoulou,
C. Moure and E. Sawaguchi for their helpful comments. Support by NATO
(R.C.G. 970037), the Spanish CICyT (PB96-0037) and the U.S. Department of
Energy, Division of Materials Sciences (contract No. DE-AC02-98CH10886) is
also acknowledged.

\newpage

\begin{figure}[htb]
\epsfig{width=0.80 \linewidth,figure=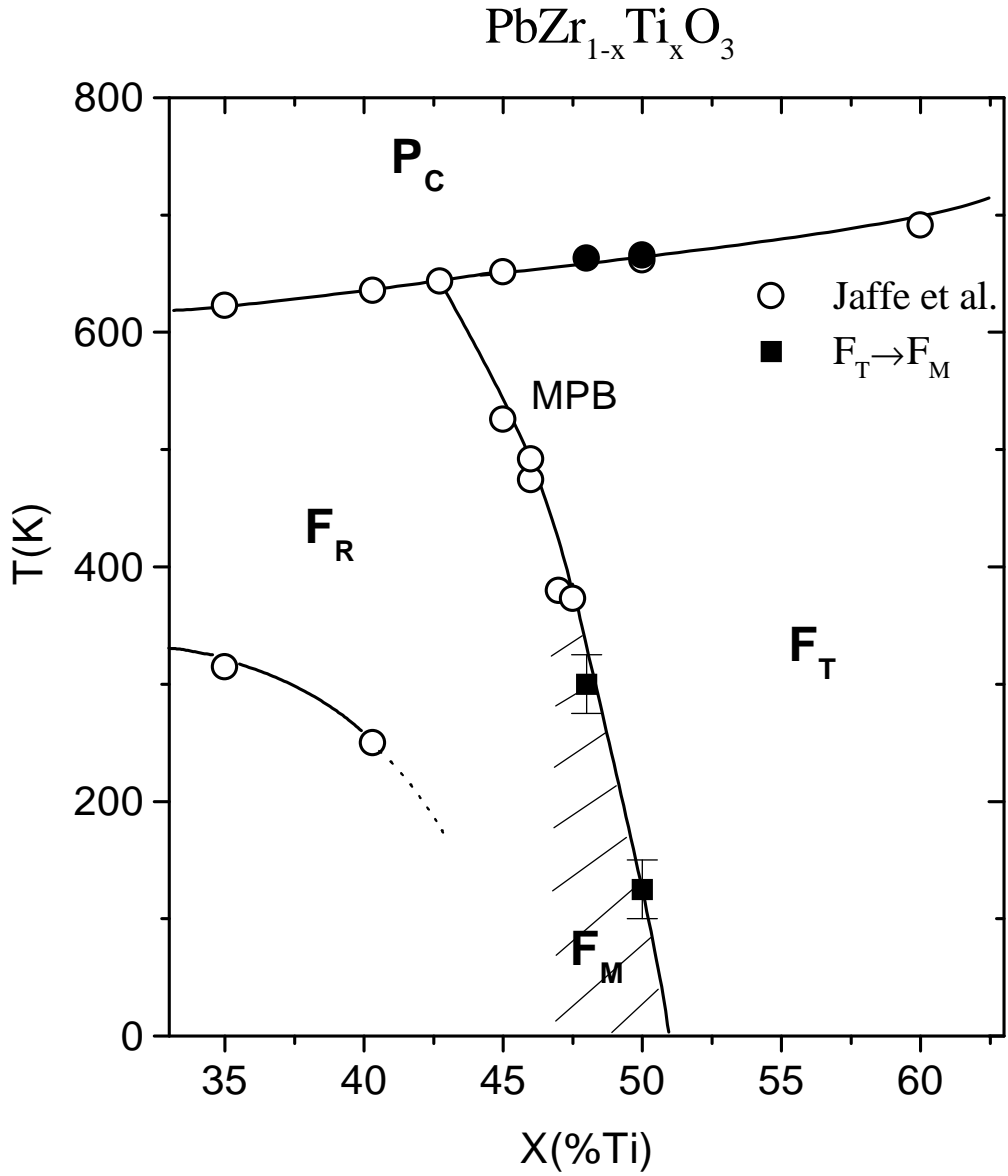} 
\vspace{0.0truecm}
\caption{Preliminary modification of the PZT phase diagram. Jaffe {\it et al.%
}'s data\protect\cite{Jaffe} are plotted as open circles. The $F_{T}-F_{M}$
and $P_{C}-F_{T}$ transition temperatures for x= 0.48 and x= 0.50 are
plotted as solid symbols. The $F_{T}-F_{M}$ transition for x= 0.50 is
reported in ref. [22].}
\label{fig1}
\end{figure}

\begin{figure}[htb]
\epsfig{width=0.80 \linewidth,figure=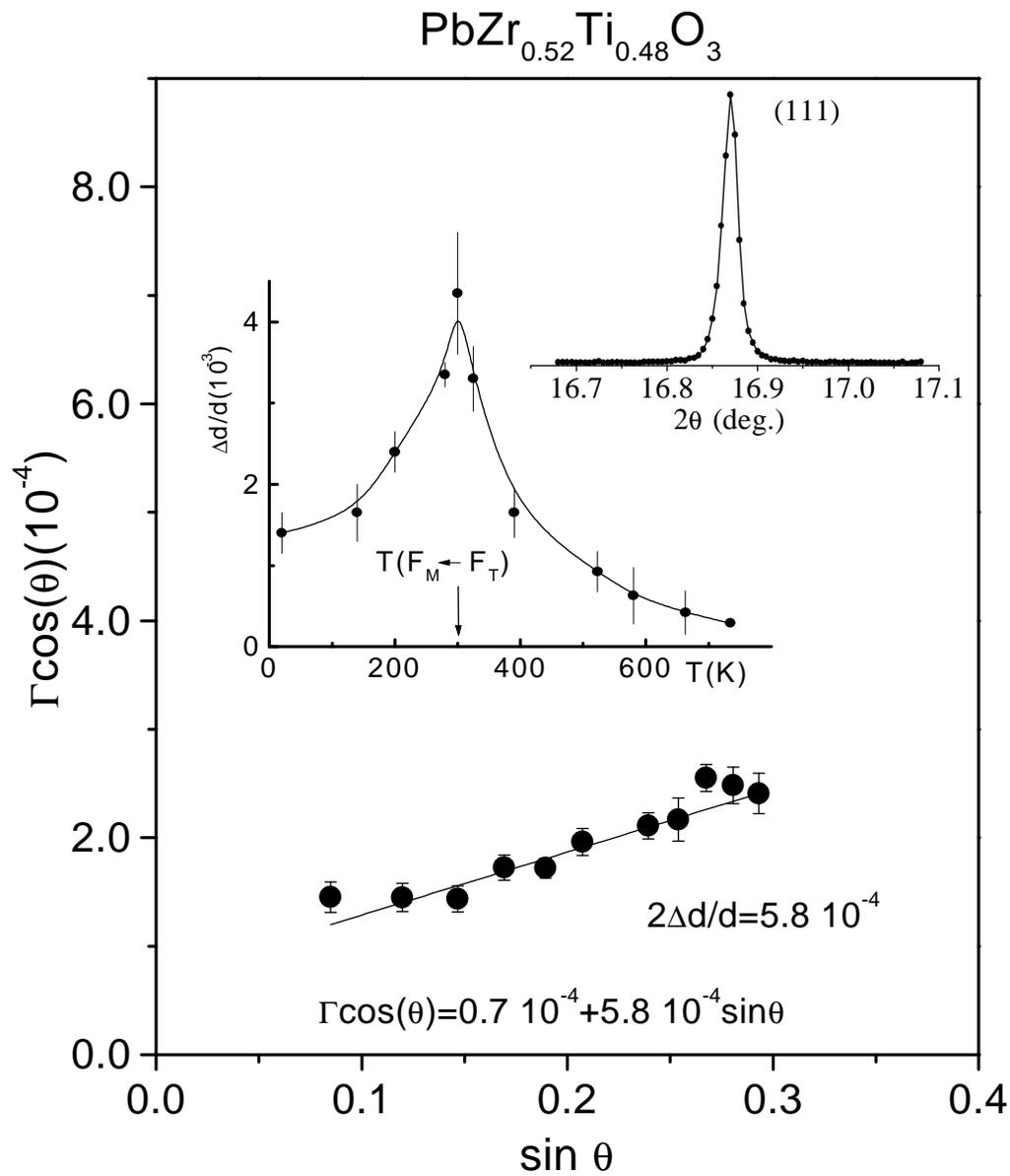} 
\vspace{0.0truecm}
\caption{The Williamson-Hall plot for PZT (x= 0.48) derived from the
measured diffraction peak widths in the cubic phase (T= 736 K). Particle
size and microstrain are estimated from a linear fit (solid line). The plot
for the (111) reflection in the cubic phase demonstrates the excellent
quality of the ceramic sample (peak width $\sim 0.02^{o}$). The plot of $%
\Delta d/d$ vs. temperature is also shown as an inset.}
\label{fig2}
\end{figure}

\begin{figure}[htb]
\epsfig{width=0.80 \linewidth,figure=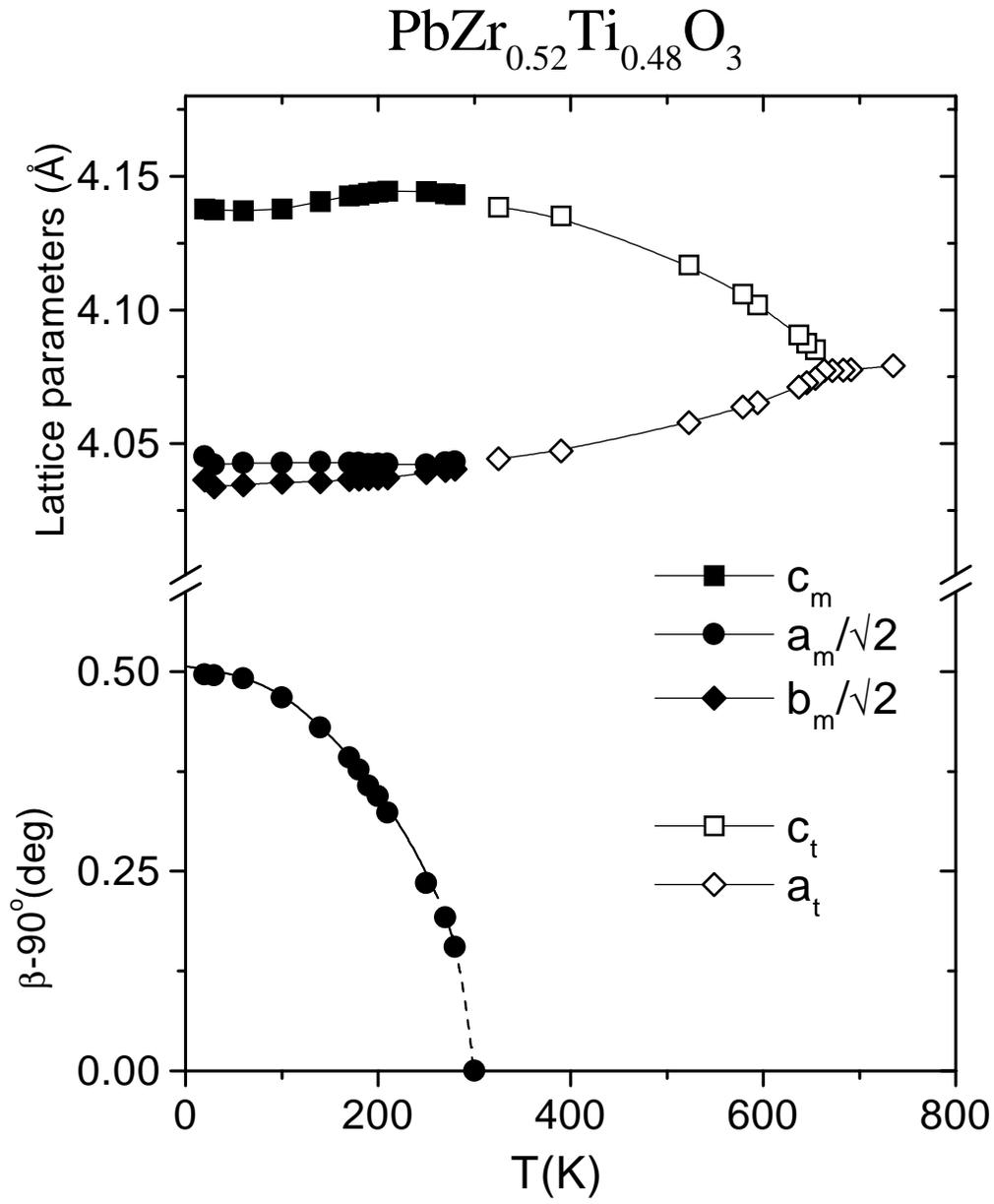} 
\vspace{0.0truecm}
\caption{Lattice parameters versus temperature for PZT (x=0.48) over the
whole range of temperatures from 20 K to 750 K showing the evolution from
the monoclinic phase to the cubic phase via the tetragonal phase.}
\label{fig3}
\end{figure}

\begin{figure}[htb]
\epsfig{width=0.80 \linewidth,figure=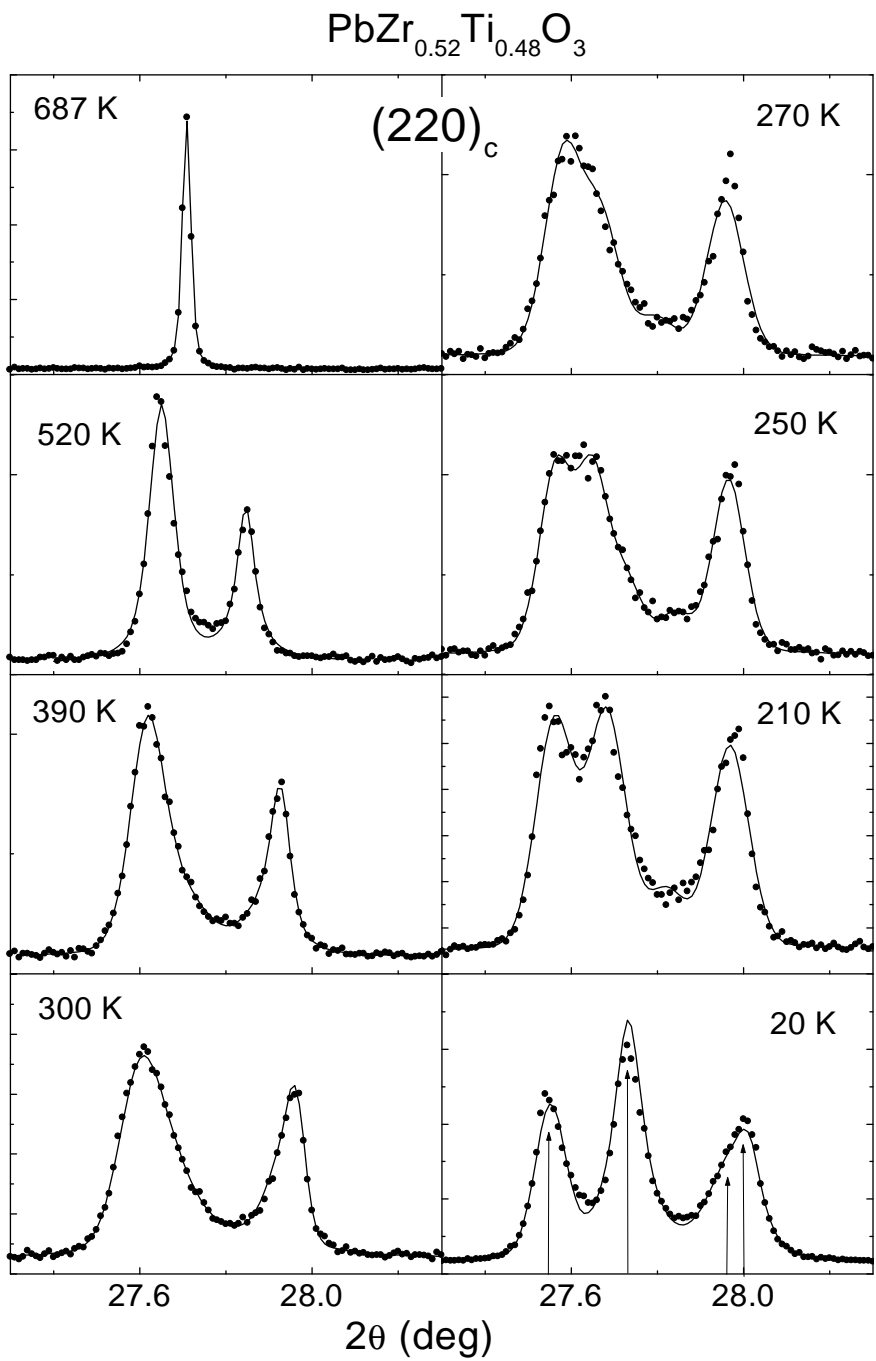} 
\vspace{0.0truecm}
\caption{Temperature evolution of the pseudo-cubic (220) peak from the cubic
(top left) to the monoclinic (bottom right) phase.}
\label{fig4}
\end{figure}

\begin{figure}[htb]
\epsfig{width=0.85 \linewidth,figure=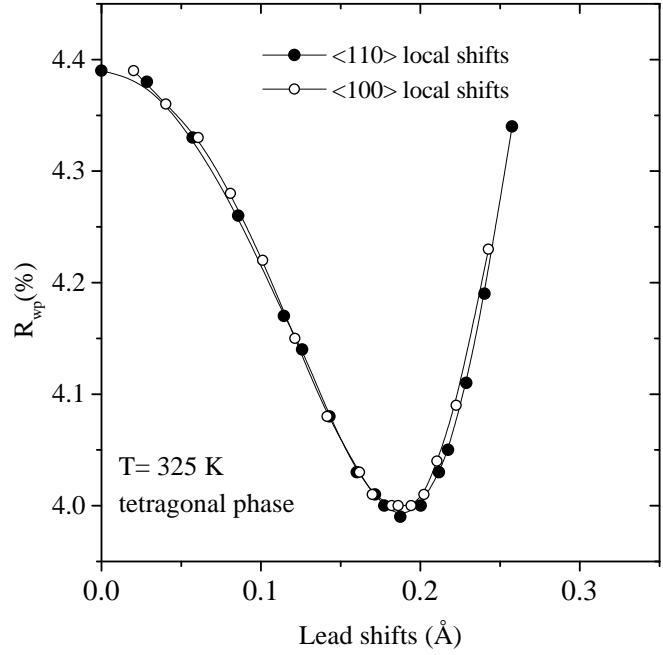} 
\vspace{0.0truecm}
\caption{Agreement factor, R$_{wp}$, as a function of Pb displacements for refinements 
with fixed values of x along tetragonal $\langle $110$\rangle $ and $\langle $100$\rangle $ directions as described in text. The well-defined minimum at x$\sim $0.19 \AA\  confirms the
the result listed in Table I for model II.}
\label{fig5}
\end{figure}

\begin{figure}[htb]
\epsfig{width=0.75 \linewidth,figure=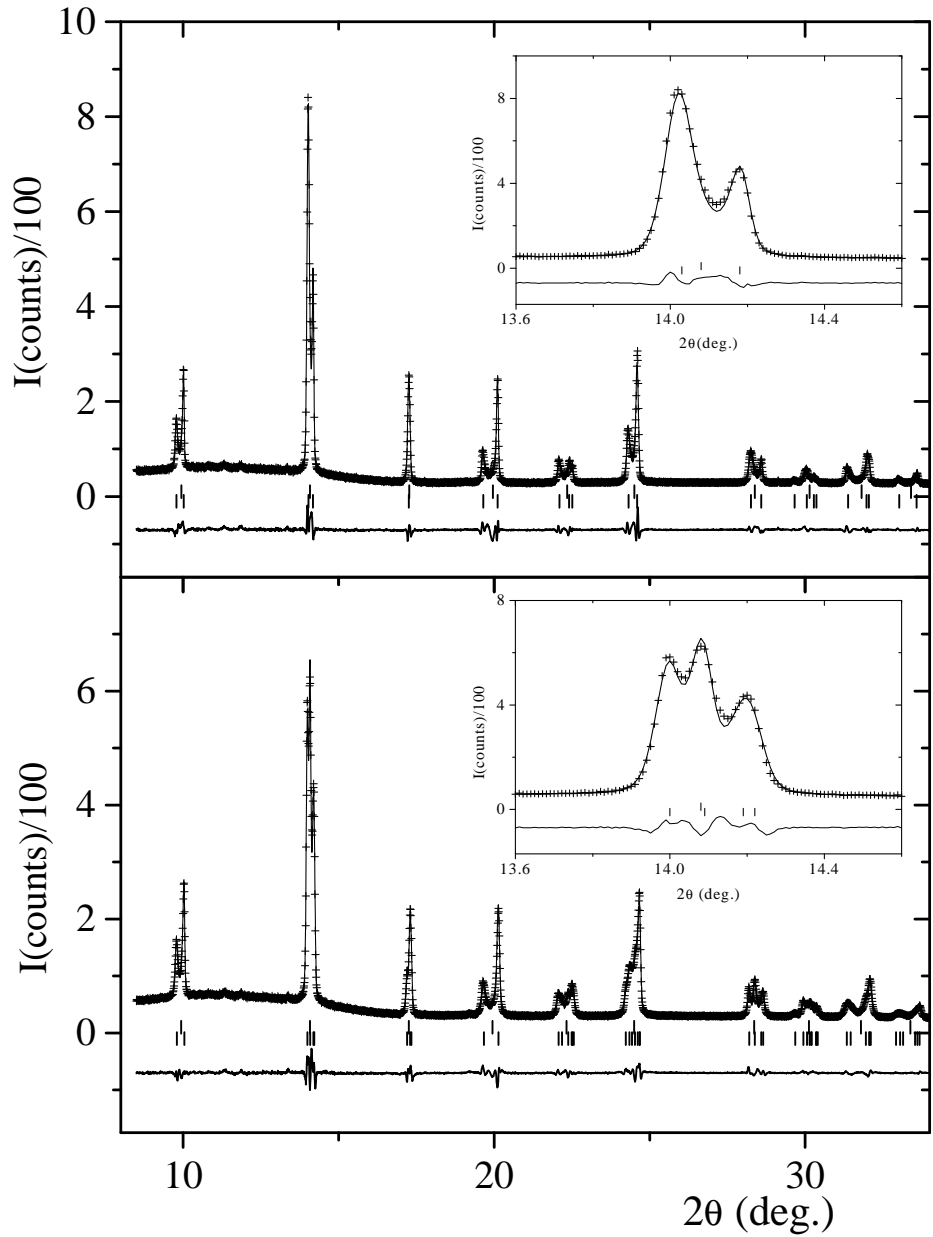} 
\vspace{0.0truecm}
\caption{Observed and calculated diffraction profiles from the Rietveld
refinement of the tetragonal (top) and monoclinic (bottom) phases of PZT (x=
0.48) at 325 and 20 K, respectively. The difference plots are shown below,
and the short vertical markers represent the peak positions (the upper set
correspond to the cubic phase as discussed in text).The insets in each
figure highlight the differences between the tetragonal and the monoclinic
phase for the pseudo-cubic (110) reflection, and illustrate the high
resolution needed in order to characterize the monoclinic phase.}
\label{fig6}
\end{figure}

\begin{figure}[htb]
\epsfig{width=0.85 \linewidth,figure=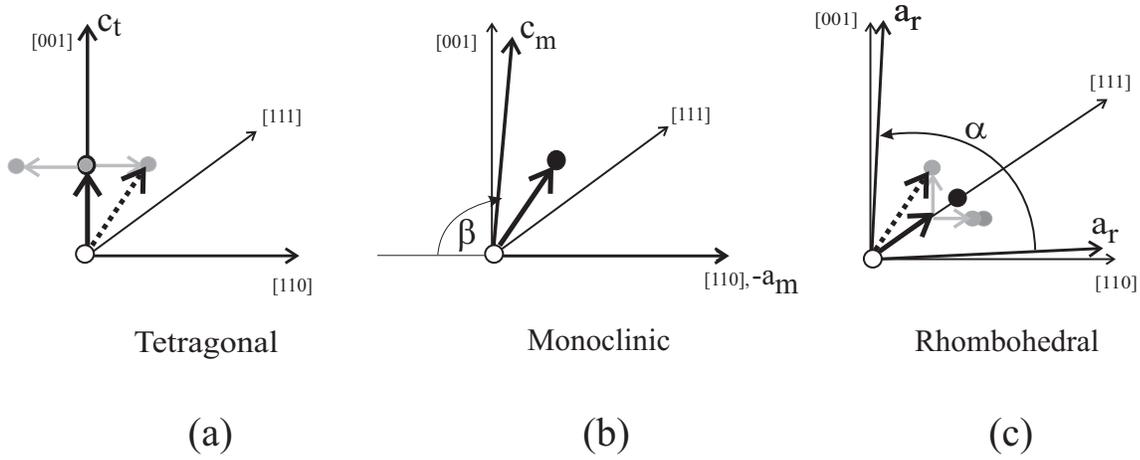} 
\vspace{0.0truecm}
\caption{Schematic illustration of the tetragonal (a), monoclinic (b) and
rhombohedral (c) distortions of the perovskite unit cell projected on the
pseudo-cubic (110) plane. The solid circles represent the observed shifts
with respect to the ideal cubic structure. The light grey circles represent
the four locally-disordered $\langle 100\rangle $ shifts in the tetragonal
structure (a) and the three locally-disordered shifts in the rhombohedral
structure (c) described by Corker {\it et al.} \protect\cite{Corker}. The
heavy dashed arrows represent the freezing-out of one of these shifts to
give the monoclinic observed structure. Note that the resultant shifts in
the rhombohedral structure can be viewed as a combination of a [111] shift
with local $\langle 100\rangle $ shifts, as indicated by the light grey
arrows.}
\label{fig7}
\end{figure}

\begin{table}[tbp]
\caption{Structure refinement results for tetragonal PbZr$_{0.52}$Ti$_{0.48}$%
O$_{3}$ at 325 K, space group P4mm, lattice parameters $a_{t}$= 4.0460(1)
\AA , $c_{t}$= 4.1394(1) \AA . Fractional occupancies, N, for all atoms
taken as unity except for Pb in model II, where N= 0.25.}
\label{Table1}
\begin{tabular}{ccccccccc}
& \multicolumn{4}{c}{Model I} & \multicolumn{4}{c}{Model II} \\ 
& \multicolumn{4}{c}{anisotropic lead temperature factors} & 
\multicolumn{4}{c}{local  $\langle $110$\rangle $ lead shifts} \\ 
\hline
& {\ x} & {\ y} & {\ z} & U(\AA $^{2}$) & {\ x} & {\ y} & {\ z} & {\ U$%
_{iso} $}(\AA $^{2}$) \\ \hline
\ \ \ \ {\ Pb\ \ \ \ \ } & 0 & 0 & 0 & U$_{11}$=0.0319(4) & 0.0328(5) & 
0.0328(5) & 0 & 0.0127(4) \\ 
&  &  &  & U$_{33}$=0.0127(4) &  &  &  &  \\ 
\ \ {Zr/Ti\ \ \ } & 0.5 & 0.5 & 0.4517(7) & U$_{iso}$=0.0052(6) & 0.5 & 0.5
& 0.4509(7) & 0.0041(6) \\ 
\ \ \ {\ O}${(1)}$\ \ \  & 0.5 & 0.5 & -0.1027(28) & U$_{iso}$=0.0061(34) & 
0.5 & 0.5 & -0.1027(28) & 0.0072(35) \\ 
\ \ \ {\ O}${(2)}$\ \ \  & 0.5 & 0 & 0.3785(24) & U$_{iso}$=0.0198(30) & 0.5
& 0 & 0.3786(24) & 0.0197(30) \\ \hline
\ R$_{wp}$ & \multicolumn{4}{c}{4.00\%} & \multicolumn{4}{c}{3.99\%} \\ 
R$_{F^{2}}$ & \multicolumn{4}{c}{6.11\%} & \multicolumn{4}{c}{6.04\%} \\ 
$\chi _{{}}^{2}$ & \multicolumn{4}{c}{$11.4$} & \multicolumn{4}{c}{$11.3$}
\end{tabular}
{\footnotesize {Agreement factors, R}}$_{{wp}}${\footnotesize {, R}}$_{{F}^{{%
2}}}${\footnotesize {\ and }}$\chi ^{2}$ {\footnotesize \ are defined in
ref.[\protect\cite{McCusker}]}.
\end{table}

\begin{table}[tbp]
\caption{Selected Zr/Ti-O and Pb-O bond lengths in the tetragonal and
monoclinic structures. Models I and II refer to the refinements with
anisotropic temperature factors and local $\langle $110$\rangle $
displacements for Pb, respectively, (see Table I). The standard errors in
the bond lengths are $\sim 0.01$ \AA .}
\label{Table2}
\begin{tabular}{cccc}
& \multicolumn{3}{c}{Bond lengths (\AA )} \\ \hline
& \multicolumn{2}{c}{tetragonal} & monoclinic \\ 
& model I & model II &  \\ \hline
Zr/Ti-O(1) & 1.85$\times 1$ & 1.85$\times 1$ & 1.87$\times 1$ \\ 
& 2.29$\times 1$ & 2.29$\times 1$ & 2.28$\times 1$ \\ 
Zr/Ti-O(2) & 2.05$\times 4$ & 2.05$\times 4$ & 2.13$\times 2$ \\ 
&  &  & 1.96$\times 2$ \\ 
Pb-O(1) & 2.89$\times 4$ & 2.90$\times 2$ & 2.90$\times 2$ \\ 
&  & 2.71$\times 1$ & 2.60$\times 1$ \\ 
Pb-O(2) & 2.56$\times 4$ & 2.67$\times 2$ & 2.64$\times 2$ \\ 
&  & 2.46$\times 2$ & 2.46$\times 2$%
\end{tabular}
\end{table}

\begin{table}[tbp]
\caption{Structure refinement results for monoclinic PbZr$_{0.52}$Ti$_{0.48}$%
O$_{3}$ at 20 K, space group Cm, lattice parameters a$_{m}$=5.72204(15)\AA ,
b$_{m}$= 5.70957(14)\AA , c$_{m}$=4.13651(14)\AA , $\protect\beta =$90.498(1)%
$^{o}$. Agreement factors R$_{wp}$= 3.26\%, \ \ R$_{F^{2}}$= 4.36\%, \ \ $%
\protect\chi ^{2}$=9.3.}
\label{Table3}
\begin{tabular}{ccccc}
& {\ x$_{m}$} & {\ y$_{m}$} & {\ z$_{m}$} & {\ U$_{iso}$}(\AA $^{2}$) \\ 
\hline
{\ Pb} & 0 & 0 & 0 & 0.0139$^{1}$ \\ 
{\ Zr/Ti} & 0.5230(6) & 0 & 0.4492(4) & 0.0011(5) \\ 
{\ O(1)} & 0.5515(23) & 0 & -0.0994(24) & 0.0035(28) \\ 
{\ O(2)} & 0.2880(18) & 0.2434(20) & 0.3729(17) & 0.0123(22)
\end{tabular}
$^{1}${\footnotesize \ For Pb, U$_{iso}$ is the equivalent isotropic value
calculated from the refined anisotropic values (U$_{11}$= 0.0253(7) \AA $%
^{2},$U$_{22}$= 0.0106(6) \AA $^{2},$U$_{33}$= 0.0059(3) \AA $^{2},$U$_{13}$%
= 0.0052(4) \AA $^{2}$)} .\newline
\end{table}

\begin{table}[tbp]
\caption{Comparison of refined values of atomic coordinates in the
monoclinic phase with the corresponding values in the tetragonal and
rhombohedral phases for both the ``ideal'' structures and those with local
shifts, as discussed in text.}
\label{Table4}
\begin{tabular}{cccccc}
& \multicolumn{2}{c}{tetragonal} & monoclinic & \multicolumn{2}{c}{
rhombohedral (ref.[\protect\cite{Amin}])} \\ 
& \multicolumn{2}{c}{x=0.48, 325 K} & x=0.48, 20K & \multicolumn{2}{c}{
x=0.40, 295 K} \\ \hline
& ideal & local shifts$^{1}$ & as refined & local shifts$^{2}$ & ideal \\ 
\hline
x$_{Zr/Ti}$ & 0.500 & 0.530 & 0.523 & 0.520 & 0.540 \\ 
z$_{Zr/Ti}$ & 0.451 & 0.451 & 0.449 & 0.420 & 0.460 \\ 
x$_{O(1)}$ & 0.500 & 0.530 & 0.551 & 0.547 & 0.567 \\ 
z$_{O(1)}$ & -0.103 & -0.103 & -0.099 & 0.093 & -0.053 \\ 
x$_{O(2)}$ & 0.250 & 0.280 & 0.288 & 0.290 & 0.310 \\ 
y$_{O(2)}$ & 0.250 & 0.250 & 0.243 & 0.257 & 0.257 \\ 
z$_{O(2)}$ & 0.379 & 0.379 & 0.373 & 0.393 & 0.433 \\ \hline
a$_{m}($\AA ) & \multicolumn{2}{c}{5.722} & 5.722 & \multicolumn{2}{c}{5.787}
\\ 
b$_{m}($\AA ) & \multicolumn{2}{c}{5.722} & 5.710 & \multicolumn{2}{c}{5.755}
\\ 
c$_{m}($\AA ) & \multicolumn{2}{c}{4.139} & 4.137 & \multicolumn{2}{c}{4.081}
\\ 
$\beta (^{o})$ & \multicolumn{2}{c}{90.0} & 90.50 & \multicolumn{2}{c}{90.45}
\end{tabular}
$^{1}${\footnotesize \ Tetragonal local shifts of (0.03,0.03,0).}\newline
$^{2}${\footnotesize \ Hexagonal local shifts of (-0.02,0.02,0).}
\end{table}


\begin{references}
\bibitem{footnote}  The [$\overline{1}\overline{1}0]$ and [1$\overline{1}0$]
directions are chosen so that the monoclinic angle $\beta >90^{o}$ to
conform with usual crystallographic convention.
\bibitem{footnote2}  The structure factor correction is defined in terms of
the anisotropic $u_{ij}$ thermal factors as $exp\{-[2\pi ^{2}(\sum
i,ju_{ij}a_{i}^{\ast }a_{j}^{\ast })]\}$, $a_{i}^{\ast }$ being the lattice
vectors of the reciprocal unit cell.
\bibitem{footnote3}  The rhombohedral unit cell can be expressed in terms of
a monoclinic one by: $a_{m}=2a_{r}\cos (\alpha /2),$ $b_{m}=2a_{r}\sin
(\alpha /2),$ $c_{m}=a_{r}$, $\beta =180^{o}-\phi $, where cos$\phi =\frac{%
1-2\sin ^{2}(\alpha /2)}{\cos (\alpha /2)}$ and $a_{r}$ and $\alpha $ are
the R3m cell parameters. Note that $a_{r}$ in ref. [37] refers to
the doubled cell.
\end{references}
\end{document}